\documentclass{aipproc}

\layoutstyle{6x9}

\begin{document}

\title
      {The world according to R{\'{e}}nyi: thermodynamics
of fractal systems}




\author{Petr Jizba}{
  address={Institute of Physics, University of Tsukuba,
  Ibaraki, 305-8571, Japan},
  email={petr@cm.ph.tsukuba.ac.jp},
}

\author{Toshihico Arimitsu}{
  address={Institute of Physics, University of Tsukuba,
Ibaraki, 305-8571, Japan},
  email={arimitsu@cm.ph.tsukuba.ac.jp}
}

\copyrightyear  {2001}

\begin{abstract}
We discuss a basic thermodynamic properties of systems
with multifractal structure. This is possible by extending the notion
of  Gibbs--Shannon's entropy into more general framework -
R{\'e}nyi's information entropy. We show a connection of R{\'e}nyi's
parameter $q$ with the multifractal singularity spectrum $f(\alpha)$ and
clarify a relationship with the Tsallis--Havrda--Charvat entropy.
Finally, we generalize Hagedorn's statistical theory and apply it to
high--energy particle collisions.
\end{abstract}

\date{\today}

\maketitle

\section{Introduction}


One of the fundamental observations of information theory is that the
most general functional form for the mean transmitted information (i.e.,
information measure) is that of R{\'e}nyi.  Although R{\'e}nyi's
information measure offers perhaps the most general and conceptually
cleanest setting for the entropy, it has not found so far as much
applicability as its Shannon's counterpart.  To clarify the
position of R{\'e}nyi's entropy in physics, we resort to systems with a
multifractal structure. Such systems are very important and highly
diverse, including phase transitions, turbulent flow of fluids,
irregularities in heartbeat, population dynamics, chemical reactions,
plasma physics, and most recently the motion of groups and clusters of
stars. We shall argue that  for the aforementioned the R{\'e}nyi
parameter is connected via a  Legendre transformation with the
multifractal singularity spectrum.  To put some flesh on bones we
generalize Hagedorn's statistical theory and subsequently apply to   a
differential cross section in high--energy scattering experiments.
More thorough investigation will be published elsewhere.


\section{R{\'e}nyi's entropy}

\subsection{Motivation}

From information theory follows that the most
general information entropy is that of
R\'{e}nyi \cite{Renyi1954}.
In discrete cases where the probability distribution
${\mathcal{P}} = \{ p_n \}$ the R\'{e}nyi entropy is defined as
\begin{displaymath}
{\mathcal{I}}_{q}({\mathcal{P}}) = \frac{1}{(1-q)}\, \log_2 \left(
\sum_{k=1}^n p_k^q \right)\, .
\end{displaymath}

On the other hand, in continuous probability 
cases a renormalization is needed - with an arbitrary precision
of measurement comes infinity of information.  If $f(x)$  is a
probability
density, say in the interval $[a,b]$ one may
define the integrated probability
\begin{displaymath}
p_{nk}= \int_{k/n}^{(k+1)/n}f(x)dx \, ,
\end{displaymath}
then \cite{Renyi1954}
\begin{equation}
{\mathcal{I}}_{q}(f) \equiv  \lim_{n\rightarrow \infty}
\left( {\mathcal{I}}_{q}({\mathcal{P}}_{nk}) - \log_2 n \right)
= \frac{1}{(1-q)}\, \log_2 \left(\int_a^b f^q(x)dx \right)\, .
\label{Ren2}
\end{equation}
Eq.\eqref{Ren2} might be generalized to any Lebesgue or
Hausdorff measurable sets \cite{Barnsley1988}. 

In the former context a natural question arises; how comes that there are
other information entropies apart from Shannon's one. 
To understand this we should go to information theory. The latter 
asserts that the amount of information received by
learning that an event of probability $p$ took place (in
bit units) is ${\mathcal{I}}(p) = - \log_2(p)$. In
general, if the possible outcomes of an experiment  are
${\cal{A}}_1, {\cal{A}}_2, \ldots, {\cal{A}}_n$ with
corresponding probabilities $p_1, p_2, \ldots, p_n$, and
${\cal{A}}_k$ conveys ${\cal{I}}_k$ bits, then the 
mean conveyed information is
\begin{displaymath}
{\cal{I}}({\cal{P}}) = \sum_{k=1}^{n}p_k {\cal{I}}_k\, .
\end{displaymath}

However, the linear averaging is only a specific case of a
more general mean! It has been recognized by
A.Kolmogorov \cite{Kolmogorov1930} and M.Nagumo \cite{Nagumo1930} 
that the most general mean
compatible with postulates of probability theory gives the entropy
\begin{displaymath}
{\cal{I}}_g ({\cal{P}}) = g^{-1}\left( \sum^n_k p_k \ g({\cal{I}}_k)
\right)\, ,
\end{displaymath}
\noindent where $g$ is an arbitrary invertible function.

Applying the postulate of additivity
of independent information one obtains only
two possible classes of $g$ \cite{Renyi1954}, namely
$g(x) = cx + d$  which implies
${\mathcal{I}}({\mathcal{P}}) = - \sum_{k=1}^n p_k \log_2 (p_k)$ (i.e.,
Shannon's information measure) and
$g(x) = c 2^{(1-q)x} + d$ which implies directly R\'{e}nyi's
information measure
${\mathcal{I}}_q({\mathcal{P}})$.

Among the basic properties of R\'{e}nyi's entropy we may mention;
positivity (${\mathcal{I}}_{q} \geq 0$), for $q\leq 1$ R\'{e}nyi's
entropy is concave but
for $q> 1$ is not pure convex nor pure concave and, in addition,  
when $q$ is continued to $1$, R\'{e}nyi's
entropy equals Shannon's one, i.e., 
$\lim_{q \rightarrow 1} {\mathcal{I}}_q = {\mathcal{I}}$.

In physics R\'{e}nyi's entropy has been 
sporadically, albeit successfully applied in various non--equilibrium
dynamical systems, e.g., fully developed turbulence ($q$ directly
relates with Reynolds number), percolating clusters ($q$
directly describes $p_c$), etc.. It also provides a
consistent mathematical setting for Tsallis--Havrda--Charvat (THC)
entropy and, as we shall see, it is a
correct entropy for (multi)fractal systems.

\subsection{Connection with Tsallis--Havrda--Charvat entropy}

THC entropy introduced originally be J.H.Havrda and F.Charvat
\cite{HC1967} and later applied to physical problems by C.Tsallis
\cite{T1987} is currently fruitfully used in many statistical systems;
3--dimensional fully developed hydrodynamic
turbulence, 2--dimensional turbulence in pure electron plasma,
Hamiltonian systems with long--range interactions, granular systems,
systems with strange non--chaotic attractors,
peculiar velocities in galactic clusters, etc.. Its form reads
\begin{displaymath}
{\cal{S}}_{q} = \frac{1}{(1-q)}\left[\sum_{k=1}^{n}(p_k)^{q}
-1\right] \, , \;\;\;\; q > 0 \, .
\end{displaymath}

Among important properties of THC entropy we can mention
positivity (${\cal{S}}_{q} \geq 0$),  concavity in ${\mathcal{P}}$,
gibbsian limit: $\lim_{q
\rightarrow 1}{\cal{S}}_{q} = {\mathcal{I}}$, and peculiar
non--extensive behavior
\begin{displaymath}
{\cal{S}}_{q}(A + B)= {\cal{S}}_{q}(A) + {\cal{S}}_{q}(B) +
(1-q){\cal{S}}_{q}(A){\cal{S}}_{q}(B)\, ,
\end{displaymath}
\noindent for two independent events $A$ and $B$.

\subsubsection{R\'{e}nyi's entropy vs.THC entropy}

To find a connection
between R{\'e}nyi and THC entropies we utilize the expansion of
$\log_2(1+x)$. Then we may write
\begin{equation}
{\cal{I}}_q = \frac{1}{(1-q)}\log_2
\left[(1-q){\cal{S}}_{q} + 1\right]
= \frac{1}{k}{\mathcal{S}}_{q} - \frac{1}{2k}
(1-q){\mathcal{S}}^2_{q} +
{\cal{O}}\left[(1-q)^2{\mathcal{S}}^3_{q}\right] \, ,
\label{TvR1}
\end{equation}
\noindent with the scale factor $k=\ln 2$. So
${\cal{I}}_q \approx {\cal{S}}_{q}$, provided
\begin{equation}
\frac{1}{|1-q|} \left[ \sum_l^n (p_l)^{q} -1 \right]^2 \ll 1 \, .
\label{cond1}
\end{equation}

Condition \eqref{cond1} is fulfilled in numerous ways.
For instance, when $q \approx 1$, $q \gg 1$, for systems with large
deviations or for rare events systems \cite{JA1}.

At this stage some comments are in order.
First of all we see from \eqref{TvR1} that
THC entropy and R{\'e}nyi's entropy are monotonic functions of
each other and, as a result, both are extremized by the same 
${\mathcal{P}}$. 
However, while R{\'e}nyi's entropy is
additive, THC entropy is not, so it appears that the
additivity property is not important for entropies required for
extremizing purposes. Thus from
thermodynamic point of view both entropies give the same predictions
(see e.g.,\cite{Ar12})!

Secondly, as we show in \cite{JA1},
R\'{e}nyi's entropy provides a consistent renormalization prescription
for a continuous THC entropy, we find that
\begin{displaymath}
{\mathcal{S}}_q (f) \equiv \lim_{n \rightarrow \infty }
\left( \frac{{\mathcal{S}}_q ({\mathcal{P}}_{nk})}{n^{(1-q)}}
- \frac{{\mathcal{S}}_q(1/n)}{n^{(1-q)}} \right) 
= \frac{1}{(1-q)} \, \int_a^b dx \, f(x)\left( f^{q-1}(x) - 1\right)\, .
\end{displaymath}

\section{ FRACTAL AND MULTIFRACTAL SYSTEMS}

\subsection{Brief introduction into (multi)fractal sets}

\subsubsection{Fractals}

Let us begin to illustrate the basic features of
fractals sets on a simple example  - triadic Koch curve (TKC).
The latter is defined iteratively in the following way:
in $0$th iteration ($n=0$) we start with a straight line -  {\em
initiator} - with length $r_{0} = a$. In the following step ($n=1$)
we raise an equilateral triangle over the middle third of initiator.
The result is {\em generator}. Its four straight line segments ($N_1 =
4$) have 
length $r_1 =
a/3 $ and total length $L[ a/3 ]
= 4a/3 $.
The construction of the Koch curve proceeds by replacing
each segment of initiator with generator, i.e., for
$n=2$, $r_2 = \left( 1/3 \right)^2 a$,
$L[\left(1/3\right)^2 a] = \left(
4/3\right)^2 a$ and $N_2 = 16$, etc. So when 
$n=k$ we have $r_k = \left( 1/3
\right)^ka$, $L[ ( 1/3) ^k a ] =
\left( 4/3\right)^k a$ and $N_k = 4^k$.
Note that the length $L$ diverges as $k \rightarrow \infty$~!

Can we define somehow a finite length for the triadic Koch curve?
The answer is yes, provided we extend the notion of
euclidean dimension. To see this let us note that for ordinary smooth
curves the approximative length is
$L[r] \sim N(r)r$, and as $r$ goes to
zero $L[r]$ approaches the finite limit - length; 
$L = \lim_{r \rightarrow 0} N(r)r$.

Generalization to any $D$--dimensional volumes is then natural:
$V = \lim_{r \rightarrow 0} N(r)r^D$. However, in order to get 
$V$ finite, the following scaling must apply
\begin{displaymath}
N(r) \sim \frac{c}{r^D} \,\, \Leftrightarrow \,\, \log N(r)
\sim c + D \log \frac{1}{r}\, ,
\end{displaymath}
\noindent and so
\begin{equation}
\lim_{r \rightarrow 0} \frac{\log N(r)}{\log \frac{1}{r}} = D\, .
\label{HB1}
\end{equation}

As the LHS of \eqref{HB1} is well defined for wider class of
sets than just usual metric spaces one may accept it as a generalized
definition of dimension. The latter is usually called the
{\em Hausdorff--Besicovitch}  or {\em fractal} dimension. 
It should be stressed that
$D$ in \eqref{HB1} is not necessarily integer - price which is paid
for the finiteness of the volume.

Thus, for instance, in the case of TKC the fractal
dimension is
\begin{displaymath}
D = \lim_{r\rightarrow 0} \frac{\log N(r)}{\log \frac{1}{r}} =
\lim_{n \rightarrow \infty} \frac{\log 4^n}{-\log \frac{a}{3^n}} =
\frac{\log 4}{ \log 3} = 1.26 \ldots \, .
\end{displaymath}

One may often write (e.g., for strictly self--similar fractals),
after $n$ iterations $N
= N_G^n $ ($N_G$ is the number of pieces of the generator)
$r = a r_G^n$ ($r_G$ is the length of the segments of the
generator). In such cases the fractal dimension follows from a 
simple analysis:
\begin{eqnarray}
&&\lim_{r \rightarrow 0} N(r)r^D = \lim_{n \rightarrow \infty}
\left( N_G r_G^D \right)^n = const. \,\,  \Rightarrow \,\, D
= \frac{\log N_G}{\log \frac{1}{r_G}}\, .
\label{frdim1}
\end{eqnarray}

Relation \eqref{frdim1} allows to recover fairly simply some 
standard results; e.g., the well known triadic Cantor dust 
($r_G = 1/3$, $N_G = 2$) has $D =
\log 2/\log 3$.

\subsubsection{Multifractals}

Multifractals are related to the study  of
a distribution of physical or other quantities on a generic support
(be it or not fractal) and thus provide a move from the geometry of
sets as such to geometric properties of distributions. An intuitive
picture about an inner structure of  multifractals is obtained by
introducing the $f(\alpha)$ spectrum \cite{Kad1}.  To elucidate the latter
let us suppose that some support  (usually a subset of a metric space)
is covered by probability of a certain phenomenon. If we
pave the support with boxes of size $l$ and denote the integrated
probability in the $i$th box as $p_i$, we may define
the scaling exponent $\alpha_i$ by 
$p_i (l) \sim l^{\alpha_i} $.
The factor $\alpha_i$ is called {\em
Lipshitz--H\"{o}lder} exponent. Counting boxes $dN(\alpha,l)$ in which $p_i$
has $\alpha_i \in (\alpha,\alpha + d\alpha)$, then
the singularity spectrum $f(\alpha)$ is defined as 
$dN(\alpha, l)= n(\alpha)l^{-f(\alpha)}d\alpha \,$ (the 
proportionality
function
$n(\alpha)$ is $l$ independent). Accordingly, we may 
view a multifractal as the ensemble of
intertwined (mono)fractals each with its own fractal dimension
$f(\alpha_i)$. It is thus suggestive to define the ``partition function''
\begin{equation}
Z(q) = \sum_i p_i^q = \int d\alpha' n(\alpha') l^{-f(\alpha')}
l^{q\alpha'} \, .
\label{pf1}
\end{equation}

In the small $l$ limit the partition function \eqref{pf1}
scales as $Z(q)\sim l^{\tau}$, where
\begin{equation}
\tau(q) = \min_{\alpha} (q \alpha - f(\alpha)), \,\,
f'(\alpha(q)) =q \, .
\label{pf3}
\end{equation}

Eq.\eqref{pf3} represents defining relations for 
the Legendre transformation.

\subsection{ R\'{e}nyi's entropy - entropy of self--similar systems}

Let us now turn to the question
whether there is any connection of R\'{e}nyi's entropy with
(multi)fractal systems.  At present it seems to us that there are
two such connections.

\subsubsection{ a) Formal connection - generalized dimensions}

\noindent Generalized dimensions are defined as:
\begin{eqnarray*}
D_q = \lim_{l\rightarrow 0} \left( \frac{1}{(q-1)} \frac{\log
Z(q)}{\log l} \right) = -\lim_{l \rightarrow 0} {\cal{I}}_q (l)\, .
\end{eqnarray*}

For example, $D_0$ is the usual fractal dimension -  dimension
of the support, $D_1$ is known as information dimension and  $D_2$ is
correlation dimension. $D_0, D_1$ and $D_2$ are usually sufficient
to describe simple fractals (e.g., strictly self similar ones). However,
in general all $D_q$ are necessary to pinpoint fractals uniquely. This
is typical 
e.g., for strange attractors \cite{HaPr1983}! The
situation is somehow analogous to statistical physics when
the whole tower of correlation function equations (BBGKY hierarchy) 
is needed to get the full information on density matrix.

\subsubsection{ b) Direct physical connection }

We will show now that from the maximal entropy (MaxEnt) point  of
view, extremizing the Gibbs--Shannon entropy on fractals is equivalent
to extremizing directly R\'{e}nyi's entropy without invoking the
underlying fractal structure explicitly.

Let us have a multifractal with a
measure $p(x)$. Shannon's entropy for the corresponding
process is ${\mathcal{I}} = - \sum p_k \log_2 p_k$.
The Billingsley theorem then states \cite{Billingsley1965} 
that there is an intimate connection between
Shannon's entropy and the Hausdorff dimension of the measure
theoretic support ${\mathcal{M}}$ of $p(x)$
(i.e., the infimum of the dimensions of all sets on which $p(x)$ 
lives). Namely,
\begin{displaymath}
d_h ({\mathcal{M}}) = - \lim_{N \rightarrow \infty}
\frac{1}{\log_2 N} \, \sum p_k \log_2 p_k \sim \frac{1}{\log_2
\varepsilon} \, \sum p_k(\varepsilon) \log_2 p_k(\varepsilon)\, ,
\end{displaymath}
\noindent with the cutoff scale $\varepsilon \sim 1/N$.
In this connection it is useful to introduce a one--parametric
family of normalized measures $\mu(q)$ (escort or zooming distributions)
\begin{displaymath}
\mu_i(q,l) = \frac{[p_i(l)]^q}{\sum_j [p_j(l)]^q}\, .
\end{displaymath}

It is important to notice that the parameter $q$  provides a
microscope for exploring different regions of the singular
measure. Indeed, for  $q >1$, $\mu(q)$ amplifies the more singular
regions of $p$, while for $q <1$, $\mu(q)$ accentuates the less singular
ones. So one may zoom into any required regions of
fractality. The corresponding ``zooming'' entropy 
$\tilde{\mathcal{I}}(q) = - \sum_k \mu_k \log_2 \mu_k$,
and $d_h$ of
the measure theoretic support of $\mu(q)$ then reads
\begin{equation}
f(q) = - \lim_{N \rightarrow \infty} \frac{1}{\log_2 N} \sum_k^N
\mu_k \log_2 \mu_k \sim \frac{1}{\log_2 \varepsilon } \, \sum_k
\mu_k (\varepsilon) \log_2 \mu_k (\varepsilon)\, .
\label{me1}
\end{equation}

In addition, the {\em average
value} of the singularity exponent $\alpha_i = \log_2 (p_i)/
\log_2 \varepsilon$ with respect to $\mu(q)$ is
\begin{equation}
\alpha(q) = \frac{\sum_k \mu_k (\varepsilon) \log_2 p_k
(\varepsilon) }{\log_2 \varepsilon}\, .
\label{me2}
\end{equation}
Eqs.\eqref{me1}  and \eqref{me2} establish a relationship
between a Hausdorff dimension $f(q)$ and an {\em average}
singularity exponent $\alpha(q)$ via functional dependence on the 
parameter
$q$. Note that $f = q\alpha - \tau$, $\alpha = d\tau /dq$ is
precisely the Legendre transformation.
Thus Shannon's entropy on a multifractal with a given
$f(\alpha)$
\begin{equation}
\left. - \sum_k p_k \log_2 p_k \right|_{f(\alpha)} = \left. - \sum_{k}
\mu_k \log_2 \mu_k \right|_{\alpha(q)}
= - q\alpha(q) \log_2 \varepsilon  + (1-q){\mathcal{I}}_q
({\mathcal{P}})\, .
\label{SR1}
\end{equation}

So as long as we fix the ``fractality'' condition, Shannon's entropy
${\mathcal{I}}$ turns out to be (up to an additive constant) R\'{e}nyi's
entropy. Thus, from MaxEnt point of view
${\mathcal{I}}|_{f(\alpha)}$ and ${\mathcal{I}}_q$ are completely
interchangeable\footnote{In \cite{JA1} we show that \eqref{Ren2}
together with  $\varepsilon \rightarrow 0$ imply that  
${\mathcal{I}}(f)|_{f(\alpha)} = (1-q){\mathcal{I}}_q(f)$. } .

\section{Generalized Hagedorn's statistical theory}

Hagedorn's statistical theory is applicable whenever
the density of quantum states grows exponentially with temperature,
i.e., when
\begin{equation}
\nu(E) \propto \exp[\beta_H E]\, .
\label{Hag1}
\end{equation}

Assumption \eqref{Hag1} applies for example to (quantized) string theory
\cite{Carlitz1972}, to cosmic string theory
\cite{MT1987Vilenkin1988}, or to high--energy particle collisions
\cite{Hagedorn1965}. 

Except a proliferation of the states near
$T_H = 1/\beta_H$, the state space acquires an approximately multifractal
structure which is exact at critical point 
\cite{Hagedorn1965,Kadanoff19821986Parisi1985}. 
This suggests that R\'{e}nyi's rather
than Gibbs entropy could better grasp vital features near $T_H$
\footnote{Actually R\'{e}nyi's statistics also modifies $T_H$. Usually 
$T_R \leq T_H$}.

\subsection{Differential cross section in high--energy scattering experiments}

Recent experiments suggest that Hagedorn's theory is not 
adequate for generic high--nergy scattering experiments. For 
instance, in $e^+ e^-$ processes  Hagedorn's description is satisfactory
provided CMS energies are small ($E<10Gev$) but it
fails at large energies. Hagedorn's approach at $E>10GeV$ predicts
an exponential decay of differential cross sections while
experiments observe a power--law behavior
\cite{DELPHI1997ZEUS1999}.

It addition, the latest applications of THC entropy
in the context of high energy collisions \cite{Beck2000}, high
energy cosmic ray physics \cite{WilkWlodarczyk2000} or a
Focker--Planck equation treatment of charmed quarks in
quark--gluon plasma \cite{WaltonRafelski2000} fit far better with
experimental data than predictions based on the 
Gibbs--Hagedorn approach.

Hagedorn's theory basically predicts that the differential cross
section in high--energy collisions should be
\begin{displaymath}
\frac{1}{\sigma} \frac{d\sigma}{dp_T} = c p_T \int_0^\infty dx
e^{-\beta \sqrt{x^2 + \mu^2}}\; ,  \; \; \mu = \sqrt{p^2_T + m^2}\, ,
\end{displaymath}
\noindent which for large $p_T$ asymptotically behaves as
\begin{displaymath}
\frac{1}{\sigma} \frac{d\sigma}{dp_T}\sim p_T^{3/2} e^{-\beta p_T}\, .
\end{displaymath}

Yet, for $ep$ high--energy collisions
the results are best fitted by a power
law \cite{DELPHI1997ZEUS1999};
\begin{equation}
\frac{1}{\sigma} \frac{d\sigma}{dp_T} \sim (1 + const\,
p_T)^{-\gamma}\, , \,\,\,\, \gamma = 5.8\pm 0.5\, .
\label{HT1}
\end{equation}

Whereas ordinary thermodynamics (and hence Hagedorn's theory itself) 
is derived by
extremizing Gibbs--Shannon entropy we extremize now ${\mathcal{I}}_q$ 
instead. Following Hagedorn \cite{Hagedorn1965} we arrive at the
probability density of transverse momenta $\rho(p_T)$ \cite{JA1};
\begin{displaymath}
\rho(p_T) \propto \, u \int_0^\infty dx \, \left(1 + (q-1)\beta 
\sqrt{x^2 + u^2 +
m_{\beta}^2}\right)^{-\frac{q}{q-1}} \, ,
\end{displaymath}
\noindent where $u = \beta p_T$ and $m_{\beta} = \beta m_0$. 
Using the fact that $\rho(p_T) \propto
\sigma^{-1}d\sigma/dp_T$ we have
\begin{equation}
\frac{1}{\sigma} \frac{d \sigma}{d p_T}
 \sim c\, \sqrt{2(q-1)} \, B\left( \frac{1}{2}, \frac{q}{q-1}
 - \frac{1}{2}\right) u^{3/2} (1 + (q-1) u)^{- \frac{q}{q-1}
+ \frac{1}{2}}\, .
\label{dif1}
\end{equation}

Formula \eqref{dif1} agrees well with the 
fit \eqref{HT1}, 
provided one suitably determines parameters $c, q, T$. This in turn may 
specify the multifractal structure of the state space.

Finally note that \eqref{dif1} has its maximum at $p_T 
= T(\frac{3}{3-q})$, on
the other hand, the experimentally measured cross sections, e.g., in
relativistic  heavy ion--collisions have maximum at roughly the same
value of $p_T \approx 180 MeV$, so $T_R \approx ( 1 - q/3) p_T  \approx
105 MeV < T_H = 158 MeV$. 

\section{Outlook and open questions}

Because of a build--in predisposition to account for self--similar systems 
R\'{e}nyi's entropy naturally aspires to be an effective tool to
describe phase transitions. It is thus a challenging task to find a closer
connection with such typical tools of critical--phenomena physics 
as are conformal and renormalization groups.

In addition, witnessing an encouraging agreement of THC non--extensive
statistics predictions  with the cosmic ray experiments
\cite{WilkWlodarczyk2000} or  heavy--ion collisions
\cite{Alberico2000Utyuzh2000} we may naturally ask whether there is a
physical grounding for R\'{e}nyi's entropy in similar (non--equilibrium)
systems.  Work along those lines is currently in progress.






\end{document}